\documentclass[10pt,pra,aps,twocolumn,superscriptaddress,nofootinbib,floatfix,longbibliography,notitlepage]{revtex4-1}
\usepackage[latin1]{inputenc}
\usepackage{amsmath,color}
\usepackage{amsfonts,natbib}
\usepackage{amssymb}
\usepackage{bm}
\usepackage{appendix}
\usepackage{breqn}
\usepackage[unicode]{hyperref}
\usepackage[normalem]{ulem}

\usepackage[percent]{overpic}

\hypersetup{
	unicode=true, 
	a4paper=true,
	plainpages=false,
	colorlinks=true,
	linkcolor=blue,
	citecolor=blue,
	filecolor=blue,
	urlcolor=blue
}
\usepackage{graphicx}

\begin{document}
\title{Topological atom optics and beyond with knotted quantum wavefunctions}
\author{Maitreyi Jayaseelan}
\thanks{maitreyi.jayaseelan@colorado.edu}
\affiliation{Department of Physics and Astronomy\char`,\, University of Rochester\char`, \,Rochester\char`,\, NY 14627\char`,\, USA}
\affiliation{Center for Coherence and Quantum Optics\char`,\,University of Rochester\char`,\, Rochester\char`,\, NY 14627\char`, \,USA}	  

\author{Joseph D. Murphree}
\affiliation{Department of Physics and Astronomy\char`,\, University of Rochester\char`, \,Rochester\char`,\, NY 14627\char`,\, USA}
\affiliation{Center for Coherence and Quantum Optics\char`,\,University of Rochester\char`,\, Rochester\char`,\, NY 14627\char`, \,USA}	

\author{Justin T. Schultz}
\affiliation{Center for Coherence and Quantum Optics\char`,\,University of Rochester\char`,\, Rochester\char`,\, NY 14627\char`, \,USA}	
\affiliation{The Institute of Optics\char`, \,University of Rochester\char`, \,Rochester\char`, \,NY 14627\char`,\, USA}

\author{Janne Ruostekoski}
\affiliation{Department of Physics\char`,\,Lancaster University\char`,\,Lancaster\char`,\,LA1 4YB\char`,\,United Kingdom}

\author{Nicholas P. Bigelow}
\thanks{nbig@pas.rochester.edu}
\affiliation{Department of Physics and Astronomy\char`,\, University of Rochester\char`, \,Rochester\char`,\, NY 14627\char`,\, USA}
\affiliation{Center for Coherence and Quantum Optics\char`,\,University of Rochester\char`,\, Rochester\char`,\, NY 14627\char`, \,USA}	
\affiliation{The Institute of Optics\char`, \,University of Rochester\char`, \,Rochester\char`, \,NY 14627\char`,\, USA}

\date{\today}
\begin{abstract}
\noindent\textbf{\LARGE Abstract}
Atom optics demonstrates optical phenomena with coherent matter waves, providing a foundational connection between light and matter. Significant advances in optics have followed the realisation of structured light fields hosting complex singularities and topologically non-trivial characteristics. However, analogous studies are still in their infancy in the field of atom optics. Here, we investigate and experimentally create knotted quantum wavefunctions in spinor Bose--Einstein condensates which display non-trivial topologies. In our work we construct coordinated orbital and spin rotations of the atomic wavefunction, engineering a variety of discrete symmetries in the combined spin and orbital degrees of freedom. The structured wavefunctions that we create map to the surface of a torus to form torus knots, M\"{o}bius strips, and a twice-linked Solomon's knot. In this paper we
demonstrate striking connections between the symmetries and underlying topologies of multicomponent atomic systems and of vector optical fields---a realization of topological atom-optics.

\end{abstract}
	\maketitle

\noindent\textbf{\LARGE Introduction}

Topology has enabled many recent advances in diverse fields of physics, from microscopic quantum systems to cosmology and elementary particle physics.  One important framework within topology is a theory of knots.  These structures have long been of interest in science and mathematics, an early example of which was Lord Kelvin's vortex atom theory of 1867 in which linked vortex strings in the aether formed the structure of atoms \cite{kelvin1867vortex_2}.   Circuit topology has led to a systematic mathematical study and classification of knots, and ideas from knot topology have opened the door to appearances of knots in a variety of different contexts in physics \cite{Kauffman_2005, stasiakIdealKnots1998}.  Important examples include the knotted solitons in the Skyrme--Faddeev model of field theory~\cite{Faddeev1997,Faddeev1999,Battye1998,Sutcliffe2007,Babaev2002} and  analogous structures in atomic systems~\cite{KawaguchiKnots2008,HallKnots2016}, in light~\cite{Sugic2021,Parmee22}, and in liquid crystals~\cite{Ackerman2017,Tai2019,AlexanderRMP2012, Smalyukh_2020}. Knot topologies have also been explored in classical and quantum fluids~\cite{Kleckner2013,KediaSuperfluids2018,Annala22}, plasmas~\cite{Smiet2015}, acoustics~\cite{Zhang2020}, biology~\cite{Sumners1990}, chemistry~\cite{Preston2021,Frisch1961}, and quantum computing~\cite{Nayak2008}. The rapid advances in engineering structured light fields~\cite{strlight_review} have allowed for preparation of singular electromagnetic field lines in the form of links and knots~\cite{Leach2004, Irvine2008, Dennis2010,Kedia2013, Larocque2018}  and the creation of complex topological features in optical polarisation.  These singular optical fields reflect the vector nature of light and  demonstrate exotic symmetry properties of the electromagnetic field including with the creation of M\"obius strips and ribbons~\cite{Bauer2015OpticalMobius, BauerMultiTwistpolarizationRibbon:2019} and knots in polarisation rotations~\cite{Pisanty2019}. In our work, these innovations are elevated by transfer to coherent atomic media. In our system, the ability to tailor the complex multicomponent wavefunction and to manipulate both internal and external degrees of freedom offers a path to forming topologically non-trivial knots and to reaching far beyond what is possible in optics. 

We report on the creation of non-trivial knotted quantum wavefunctions of a quantum degenerate spinor Bose gas.  By coupling the internal symmetries of the spinor wavefunction with its external orbital angular momenta (OAM) using Raman laser fields, we cause the wavefunction to display \emph{spin-orbit invariance} in coordinated rotations of the spinor symmetry and phase. We create spin-orbit invariant wavefunctions within the spin-1 and spin-2 hyperfine manifolds of an atomic $^{87}$Rb spinor Bose--Einstein condensate (BEC). In many optical and fluid investigations, knots are formed as real space objects of singular lines~\cite{Leach2004, Dennis2010}. In contrast, our knotted matter wave structures appear in their parameter space and therefore do not undergo vortex reconnections, providing a close analogy with condensed matter and field theory models, where the knots are formed in the mappings between the order-parameter space and real space~\cite{Faddeev1997,Battye1998,Babaev2002,HallKnots2016,Ackerman2017}. Additionally, because our knotted structures are imprinted on dilute atomic clouds, they are long-lived and retain their characteristics as the cloud ballistically expands outside a trap. The approach is versatile, requiring only different configurations of polarization and optical orbital angular momenta to realise a variety of knots and links in the quantum atomic wavefunction.  Our results will enable emulation of more general optical phenomena in atomic systems~\cite{ARamanWaveplate, SingularAtomOptics} which offer exceptional advantages---including tunable interactions and the availability of higher spin manifolds---unique to the ultracold atomic platform.

\noindent\textbf{\LARGE Results}

\noindent\textbf{Spin-orbit invariant wavefunctions}.
A simple dilute-gas scalar BEC can be well described by a macroscopic wave function characterized by a spatially dependent amplitude and phase. If the condensate has internal degrees of freedom, such as a spin-$F$ spinor BEC, then a spin-$F$ macroscopic wavefunction can be described by a multicomponent spinor $\zeta(\mathbf{r}) = (\zeta_{F}, \zeta_{F-1},...,\zeta_{-F})^T$, with $\zeta_{m_F}$ representing each $|F, m_F\rangle$ state. The spinor condensate can exist in a variety of phases characterized by different sets of relationships between $\zeta_{m_F}$ terms \cite{KawaguchiUedaSpinorBoseEinsteinCondensates}.  If we consider transformations
that include rotations of the internal spin, $\hat{F}_z$, and the macroscopic wavefunction orbital angular momentum, $\hat{L}_z$,  we can transform the wavefunction between states in a given magnetic phase. For spinor BECs, the magnetic phases that are stable solutions of the nonlinear mean-field dynamics display non-trivial discrete symmetries (see Supplementary Figure 1)~\cite{KawaguchiUedaSpinorBoseEinsteinCondensates}. These symmetries reveal themselves under spin rotation around an $n$-fold internal symmetry axis through an angle $2\pi m/n$ (for integer $m$) that, together with the rotation of the global phase, leave the wavefunction unchanged. Consider the following transformation of an $n$-fold-symmetric atomic spinor $\zeta_0$ as we traverse a closed loop in space:
\begin{align}
    {\zeta} = e^{i(j_{\lambda}-\lambda \hat{F}_z)\phi}\zeta_0,
\label{eq:spingaugesymm}
\end{align} 
where $\phi$ denotes the azimuthal angle. The single-valuedness of the wavefunction constrains the possible parameter values $(\lambda, j_{\lambda})$ for the rotation of the spin state and global phase, respectively. For a spinless scalar condensate,  $j_{\lambda}$ can only take integer values reflecting the quantization of angular momentum of the condensate, and the solutions admit topological defects such as quantized vortex lines. In the spinor case, for coordinated rotations of the spin state and the orbital part, $\lambda$ and $j_{\lambda}$ can be fractional, while keeping the entire wavefunction single-valued. This coupled spin-orbit invariance represents the symmetry that, in spinor BECs, permits a rich phenomenology of fractional vorticity~\cite{KawaguchiUedaSpinorBoseEinsteinCondensates, SemenoffZhou2007, magnusBiaxial2016}.

Consider the case where two spin states, $|F,m_F\rangle$ and $|F,m_F'\rangle$, are populated with angular momenta $\ell$ and $\ell'$, respectively. Then  $j_{\lambda} - \lambda m_F = \ell$ and $j_{\lambda} - \lambda m_F' = \ell'$, and 
\begin{align}
    \lambda  = \frac{\ell'-\ell}{m_F-m_F'}, \quad j_{\lambda}  = \frac{m_F\ell'-m_F'\ell}{m_F-m_F'}.
\label{eq:generalformslambdajlambda}
\end{align} These expressions, along with specific values for the spinors $\zeta_0$ in Eq.~\ref{eq:spingaugesymm} will be used to compute the different spin-orbit invariant wavefunctions $\zeta$ that we will discuss in later sections of this work. Different allowed combinations of $\lambda$ and $j_{\lambda}$, and of  $m_F$ and $m_F'$, characterize a complex phenomenology of solutions including defects and knots. 
Note that  $\lambda$ and $j_{\lambda}$ also determine the spin and mass flows in the condensate.

\noindent{\textbf{Torus knot topology.
    }}
The fractional spin rotations of the spin-orbit invariant states are associated with a non-trivial topology. The two coordinates $\phi$ (the azimuthal angle) and $\tilde{\phi}$ (the angle of spin rotation) each parameterize a circle $S^1 = \{ e^{i \theta}: \theta\in [0,2\pi)\}$. Together they represent the parameterization of the torus  $T^2 = S^1\times S^1$, points on the surface of which are specified by the pair $(e^{i\tilde{\phi}},\, e^{i\phi})$.  In this case, the single-valuedness requirement imposes  $\tilde{\phi} = \lambda\phi$, where $\lambda = m/n$ is the ratio of the number $m$ of fractional spinor rotations through angle $2\pi/n$ of the wavefunction around an $n$-fold symmetry axis on a full azimuthal traversal. This in turn defines paths on the surface of the torus, through the coordinates $(e^{i m\phi},\, e^{i n\phi})$, that traverse the meridional direction on the torus $m$ times and the longitudinal direction $n$ times. Here, $\lambda$ is a rational fraction, and this path is a closed space curve which defines the torus knot $K_{m,n}$.  Notice that the torus knots $K_{m,n}$ are topologically equivalent to the knots $K_{n, m}$, since the choice of longitudinal or meridional coordinate can be reversed, while the knot $K_{m,-n}$ is the mirror image of the knot.

Using the torus representation we can identify distinct wavefunctions with linked and knotted topologies differing in their spin and OAM configurations. Specifically, when $(m,n)$ are coprime, the knot $K_{m,n}$ consists of a single path on the torus. This path may or may not be truly knotted; indeed some of the simplest torus knots $K_{1, n}$, known as unknots or trivial knots, are topologically equivalent to a circle. Non-trivial knots are realised for $m,\,n\neq1$.  When $(m,n)$ are not co-prime, the pair of coordinates instead  describes a torus link, which consists of $d$-many possibly interlinked torus knots $K_{m/d, n/d}$ where $d=\text{gcd}(m,n)$ \cite{adams2004knot,Chiara2016,Hatcher:2000, rolfsen2003knots}. 

\begin{figure*}[t]
\begin{flushleft}
\begin{overpic}[width = \textwidth, tics=5, unit=1pt, grid = false, trim=0 0 0 0, clip]{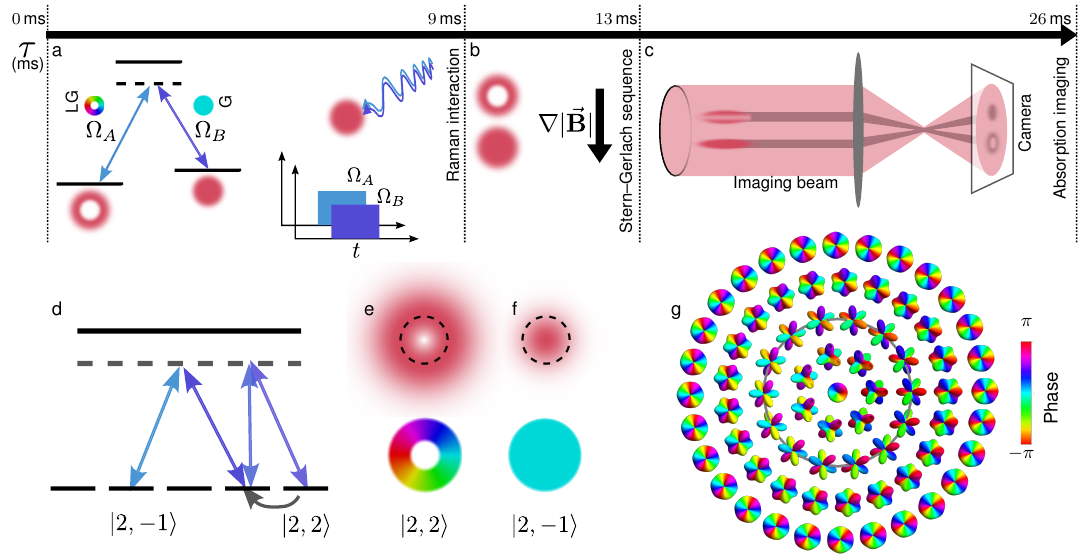}
\end{overpic}
  \caption{\textbf{Creating and detecting knotted quantum wavefunctions.} We begin with a spin-pure, spin-polarized condensate in a magnetic trap. The black arrow depicts the experimental time $\tau$. The cloud is released from the trap  at $\tau = 0$. (a) We create spin-orbit invariant wavefunctions with a coherent Raman process using optical fields with Rabi frequencies $\Omega_A$ and  $\Omega_B$, and Gaussian (G) and Laguerre--Gaussian (LG) spatial modes. The process transfers a ring-shaped region of the cloud to the final spin state and imprints an azimuthal relative phase between the spinor components, leaving a non-rotating core in the initial state. (b) We measure state populations with a time-of-flight Stern--Gerlach process using an inhomogeneous magnetic field (depicted by $\nabla|\vec{\mathbf{B}}|$). (c) An absorption image of the cloud reveals spatially resolved spin state population. (d) Multipulse Raman sequences with optical and rf fields allow more complex couplings (see Methods). We show an example sequence to create a spin-orbit invariant state where the spin state $|2, 2\rangle$ has an orbital angular momentum $\ell =1$ and a donut-shaped intensity profile as shown in (e), while the state $|2,-1\rangle$ is a non-rotating Gaussian core as shown in (f). (g) The profile of the order-parameter symmetry across the cloud, illustrated through the spherical harmonics (see Methods), can reveal the spin-orbit invariance due to the coupling between the spin and orbital angular momentum. 
  The color bar depicts phase in panels (a), (e), and (g).
      }
  \label{fig:experimentalSequence}
  \end{flushleft}
\end{figure*}

\noindent\textbf{
Experimental system: creating spin-orbit invariant states}. Our principal results are knotted wavefunctions and M\"{o}bius bands within specific spinor magnetic phases: the spin-1 polar and spin-2 cyclic and biaxial nematic phases (Supplementary Note 1) that provide different discrete internal symmetries. In the laboratory we create selected, knotted spin-orbit invariant states in the spin-1 or spin-2 electronic ground state manifolds of a rubidium spinor BEC. Both manifolds support states with unique non-trivial topologies, and the higher-order symmetries of the spin-2 manifold enable us to realize particularly complex topologies.

Experimentally creating knotted wavefunctions in the atomic cloud requires tailoring the spin state populations $|\zeta_{m_F}|^2$ and their spatially varying relative phases to control local spin state orientation and to thereby realize specific coupled symmetries.
We use a coherent optical Raman imprinting technique to engineer a target atomic wavefunction  starting from a pure $|F,m_F\rangle$ spin state (Fig.~\ref{fig:experimentalSequence}). In its simplest form, this method provides amplitude and phase controlled two-photon coupling between two states $|F,m_F\rangle$ and $|F,m_F'\rangle$ in an effective three-level $\Lambda$ system \cite{Wright:2008, Wright:2009}.  The coupling can be described by the unitary evolution operator
    \begin{align}\label{eq:UanalyticRaman}
        U(t) &= e^{i\Omega t/2} \exp\left(i \frac{\Omega t}{2} \vec{n}\cdot\vec{\sigma}\right),
    \end{align} where $\vec{n} = (\sin 2\alpha \cos \phi, \sin 2\alpha \sin \phi, \cos 2\alpha)^T$, $\vec{\sigma} = (\sigma_x, \sigma_y, \sigma_z)$ is the vector of Pauli matrices, $\phi$ is the relative phase between the Raman fields,  
    and the parameters $\Omega$ and $\alpha$ are related to the total and relative intensities of the optical fields (See Methods) \cite{schultzcreatingfbb_2016,SchultzRamanFingerprints2016}.

When one of the Raman beams has a Laguerre--Gaussian mode, this beam carries OAM. If the other Raman beam is Gaussian, a spatially varying population transfer takes place leaving a central core of atoms in the initial state $|F,m_F\rangle$ while transferring a ring-shaped population to $|F,m_F'\rangle$. The OAM of the Laguerre--Gaussian beam is also imprinted on the transferred population as an azimuthal phase twist. The handedness of the twist is determined by the beam polarizations. The Raman process therefore couples spin and OAM. Multipulse Raman sequences combine with coherent rf population transfer to generalize the two-state Raman coupling to control the spin populations, the relative phases, and the OAM, $|\ell'|$,
 of multiple $m_F'$ states. We thereby create magnetic (spinor) states and phases with specific discrete symmetries including fractional spin rotations.

\begin{figure}[t]
  \centering
  \includegraphics[trim={0cm 0cm 0cm 0cm},clip, width=\linewidth]{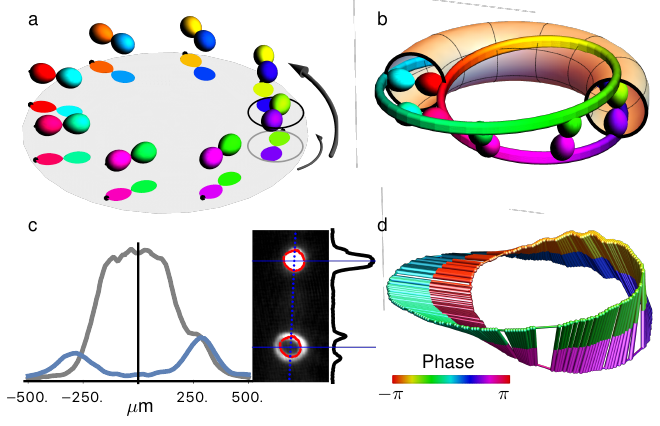}
  \caption{\textbf{Spin-orbit invariant polar phase wavefunction $\zeta^{\text{P}}$.} (a) Spherical harmonic representation of  $\zeta^{\text{P}}$  shows the coupled transformation of spin rotation angle and external phase on an azimuthal traversal of the wavefunction: a rotation of the spherical harmonics by $\pi$ is accompanied by a corresponding transformation of the overall phase by $\pi$, leaving the wavefunction single-valued. The 3D spherical harmonics are projected onto a plane (grey disk) to represent the topology of the wavefunction in 2D. (b) Torus knot topology associated with the lobe-tip path of the spherical harmonics is seen through a mapping (see Methods) from the 2D space onto a 3D torus. (c) Experimental absorption image and a lineout through the centers of the cloud spin components showing a donut-shaped intensity profile in $|1,1\rangle$ and a Gaussian core in $|1,-1\rangle$  shows a realization of $\zeta^{\text{P}}$. The red trace overlaid on the absorption image of the atomic cloud indicates regions where the atomic populations are within $1\%$ of the ideal polar phase. (d) Reconstruction of the experimentally realised M\"{o}bius band formed  by the nematic axes, showing a single half-twist of the surface. 
  The color bar depicts phase in panels (a), (b), and (d).
  }
  \label{fig:1}
\end{figure}

\noindent\textbf{Polar phase: M\"{o}bius strip topologies}. 
We consider a prototype spin-1 polar phase wavefunction $\zeta^{\text{P}}_0 \equiv \frac{1}{\sqrt{2}}\begin{pmatrix}
 1 & 0 & 1
\end{pmatrix}^T$. 
If $\ell$ and $\ell'$ are the orbital angular momenta of  $|1,1\rangle$ and $|1,-1\rangle$, from Eq.~\ref{eq:generalformslambdajlambda} 
$\lambda  = (\ell'-\ell)/2\label{eq:spin1:lambda}$ and  $j_{\lambda} =  (\ell'+\ell)/2\label{eq:spin1:jlambda}$. With $\ell = -1$ and $\ell' = 0$, we have a state with $(\lambda = 1/2,\, j_{\lambda} = -1/2)$ using
$\zeta_0$ = $\zeta_0^{\text{P}}$ in Eq.~\ref{eq:spingaugesymm}:
\begin{align}
 \zeta^{\text{P}} &=\frac{1}{\sqrt{2}}\begin{pmatrix}
 e^{-i \phi} \\ 0 \\ 1
\end{pmatrix} \label{eq:PolarHQVLemon}.
\end{align}  
We describe the specific Raman configuration that creates this atomic wavefunction in the Methods section. This wavefunction is invariant under the coupled transformation of the spin-rotation angle $\tilde{\phi} = \lambda\phi$ and external phase $\varphi = j_{\lambda}\phi$ as  $(\varphi,\,\tilde{\phi})\mapsto (\varphi-\pi,\,\tilde{\phi}+\pi)$. This symmetry is a signature of  half-quantum  vortices that have been observed in superconductors~\cite{Kirtley1996,Jang11}, superfluid $^3$He~\cite{AuttiHQVinHe3:2016}, and atomic BECs~\cite{Seo2015,Xiao21}.

Fig.~\ref{fig:1}(a) and (b) show the spherical harmonic representation (see Methods) of $\zeta^{\text{P}}$, highlighting the coupling between the spinor (through the orientation of the spherical harmonics) and the phase (through the color of the lobes). The experimental absorption image and lineout through the centers of the spin components of $\zeta^{\text{P}}$ is shown in Fig.~\ref{fig:1}(c).
The order-parameter alignment is given by the nematic axis (Supplementary Note 1).
The M\"{o}bius strip topology of the wavefunction is visualized using a construction (see Methods) which maps the spin-orbit invariant structures from the 2D space of the physical wavefunction---seen in Fig~\ref{fig:1}(a), onto a torus in 3D---seen in Fig~\ref{fig:1}(b). The associated structures are traced as the lobe-tip path of the two lobes of the spherical harmonics, and are naturally colored by the phase $j_{\lambda}\phi$, providing a simultaneous representation of both $\lambda$ and $j_{\lambda}$. The topology of $\zeta^{\text{P}}$ is that of the torus (un)knot $K_{1,2}$, which forms the edges of a M\"{o}bius band surface with a single half-twist (Fig.~\ref{fig:1}(d)). The nematic axes form a continuous surface bounded by the lobe-tip path. Starting from one of the lobe tips (say green at $\phi=0$) and making a single $2\pi$ traversal, we find that the path connects with the opposite lobe (purple), necessitating a further $2\pi$ traversal to complete the return to the original point. This is precisely the behavior of a M\"{o}bius band surface and its edge: a point on the edge of a M\"{o}bius band must complete a $4\pi$ traversal to return to itself, while the band surface itself is continuous.

For general values of $\ell$ and $\ell'$ the associated topological structures are M\"{o}bius strips with $\ell'-\ell$ half-twists, with edges of torus knots or links $K_{\ell'-\ell, 2}$. The Hopf link, with two singly linked loops, may be realised as the knot $K_{2,2}$ in a configuration with $(\ell = -2,\, \ell' = 0)$.

\begin{figure*}[t]
  \centering
  \includegraphics[width=\linewidth]{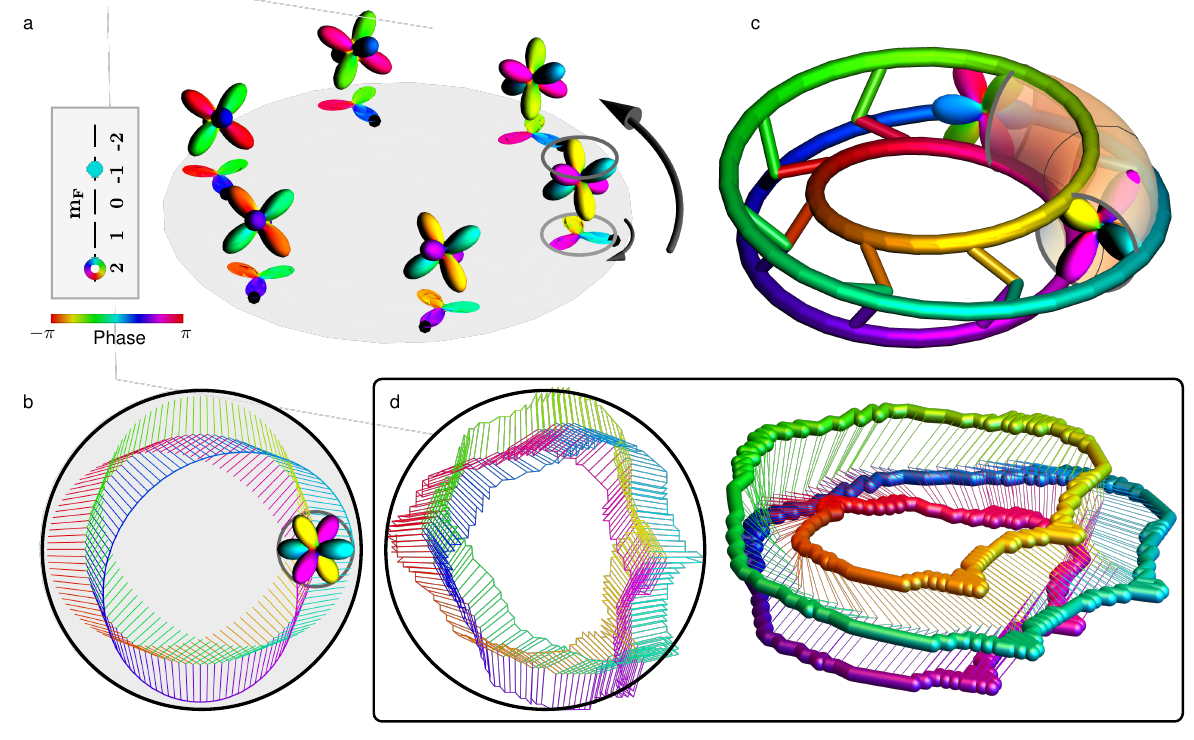}
  \caption{\textbf{Knotting states of three-fold symmetry.} (a) The spin-orbit invariant wavefunction $\zeta^{\text{C}} = \begin{pmatrix}
  \zeta_2 e^{i\phi} &  0 & 0 & \zeta_{-1} & 0 
  \end{pmatrix}^T$ shows a rotation of the spherical harmonics  by $-2\pi/3$ on a full azimuthal traversal while the overall phase changes by $2\pi/3$. In the inset box $m_F$ denotes the magnetic sublevels of the hyperfine states. We depict the populations and phases of the wavefunction components $|F, m_F\rangle$, showing the azimuthal phase of the component in $|2,2\rangle$ and the uniform phase of the component in $|2,-1\rangle$.
   (b) We show a head-on view of the three-fold symmetric spin alignment in terms of spherical harmonics using lines depicting the alignment measure  for the cyclic phase wavefunction. (c) $K_{-1,3}$ is visualized in 3D following a mapping of the coupled rotation of spin and orbital angular momentum onto the 3D torus. (d) The experimentally realized atomic wavefunction shows the rotation of spin state on an azimuthal traversal. We show a 3D reconstruction of the $K_{-1,3}$ torus knot from the experimental data.
  The color bar depicts phase in all panels.
  }
  \label{fig:2}
\end{figure*}

 A spin-$F$ atomic system can possess multipole components up to $2F$, with complex angular distributions reflected in the symmetries of the wavefunction's spherical harmonic representation. 
Spin-1 wavefunction combines a discrete two-fold symmetry with a condensate phase, resulting in the polar order-parameter symmetry of the unoriented axis and the $2\pi$ phase. Discrete polytope symmetries  can be found in higher-spin systems that support more complex topological structures, as experimentally illustrated for vortex creation~\cite{Xiao22}. 

\noindent\textbf{
Trefoil knots in the cyclic phase.}
The wavefunction of spin-2 cyclic magnetic phase $\zeta^{\text{C}}_0
\equiv\frac{1}{\sqrt{3}}\begin{pmatrix}
1 & 0 & 0 & \sqrt{2} & 0
\end{pmatrix}^T$ combines a three-fold internal symmetry under spin state rotation about the atomic quantization axis with the condensate phase. If $\ell$ and $\ell'$ are the orbital angular momenta of spin states $|2,2\rangle$ and $|2,-1\rangle$, Eq.~\ref{eq:generalformslambdajlambda} gives $\lambda  = (\ell'-\ell)/3$ and $j_{\lambda} =  (2\ell'+\ell)/3 $. The associated knot structures are $K_{\ell'-\ell,3}$. With $\ell = 1$ and $\ell' = 0$ we create a state with $(\lambda = -1/3,\,\,j_{\lambda}= 1/3)$ using $\zeta_0 = \zeta^{\text{C}}_0$ in Eq.~\ref{eq:spingaugesymm}:
\begin{align}
 \zeta^{\text{C}} & =\frac{1}{\sqrt{3}} 
 \begin{pmatrix}
 e^{i\phi} \\ 0 \\ 0 \\ \sqrt{2}\\ 0
 \end{pmatrix} \label{eq:cyclicvortex}.
\end{align}
 This state is invariant under a transformation of the spin state rotation angle $\tilde{\phi}=\lambda\phi$ and phase $\varphi = j_{\lambda}\phi$, as $(\varphi,\tilde{\phi})\mapsto (\varphi+2\pi/3, \tilde{\phi}-2\pi/3)$.

Figure~\ref{fig:2}(a) shows a spherical harmonic representation of this spin-orbit invariant state, and the lobes of the spin alignment are tracked in Fig.~\ref{fig:2}(b). Figure~\ref{fig:2}(c) shows the associated torus knot structure $K_{-1,3}$. The experimental realization of this cyclic phase wavefunction is shown in Figure~\ref{fig:2}(d) (see details in the Methods section).

\begin{figure}[t]
  \centering

  \includegraphics[trim={0.0cm 0cm 0cm 0cm},clip, width=\linewidth]{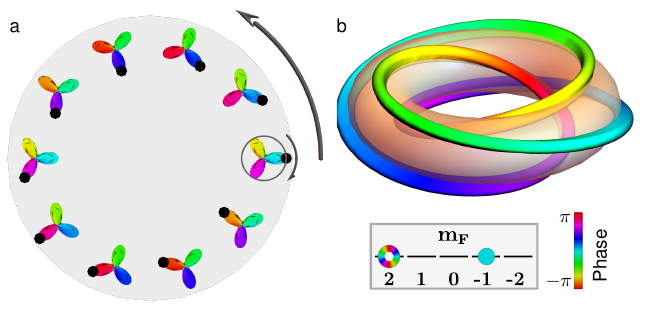}
  \caption{\textbf{Knotting true knots: the trefoil knot.} The cyclic phase also hosts more complex knots. (a, b) We show a cyclic phase wavefunction with $\ell = 2$ and $\ell' = 0$ the orbital angular momentum of spin states $|2,2\rangle$ and $|2,-1\rangle$. This wavefunction has the topology of a trefoil knot.
   The trefoil knot is the simplest non-trivial knot. (a) The planar projections of the spherical harmonic lobes show a spin rotation by $-2\pi/3$ accompanied by a coordinated transformation of the  overall phase. (b) The associated knot structure, represented on the 3D torus by tracking the spherical harmonic lobes, is the trefoil knot.  In the inset box $m_F$ denotes the magnetic sublevels of the hyperfine states. We depict the populations and phases of the wavefunction components $|F, m_F\rangle$, showing the $\ell = 2$ azimuthal phase of the component in $|2,2\rangle$ and the uniform phase of the component in $|2,-1\rangle$.
  The color bar depicts phase in both panels.
  }
  \label{fig:333}
\end{figure}

As a result of the higher-order symmetry of the wavefunction, the spin-2 cyclic phase also hosts truly knotted structures that cannot be untangled to produce the simple loop or unknot without cutting the strands. The simplest of these true knots is the trefoil knot, which is of fundamental interest in knot theory. The trefoil knot possesses a definite handedness which makes the knot and its mirror image distinct. The two distinct trefoil knots, $K_{2,3}$ and its mirror image $K_{-2,3}$, are realised when $|\ell'-\ell|=2$. Fig.~\ref{fig:333}(a) shows a cyclic phase wavefunction hosting a trefoil knot: the state undergoes a rotation $\lambda = -2/3$ that combines the
three-fold symmetry of the internal state with a coordinated change of phase. The associated torus knot shown in Fig.~\ref{fig:333}(b) is a true knot.

\begin{figure*}[t]
  \centering

  \includegraphics[trim={0cm 0cm 0cm 0cm},clip,width=\linewidth]{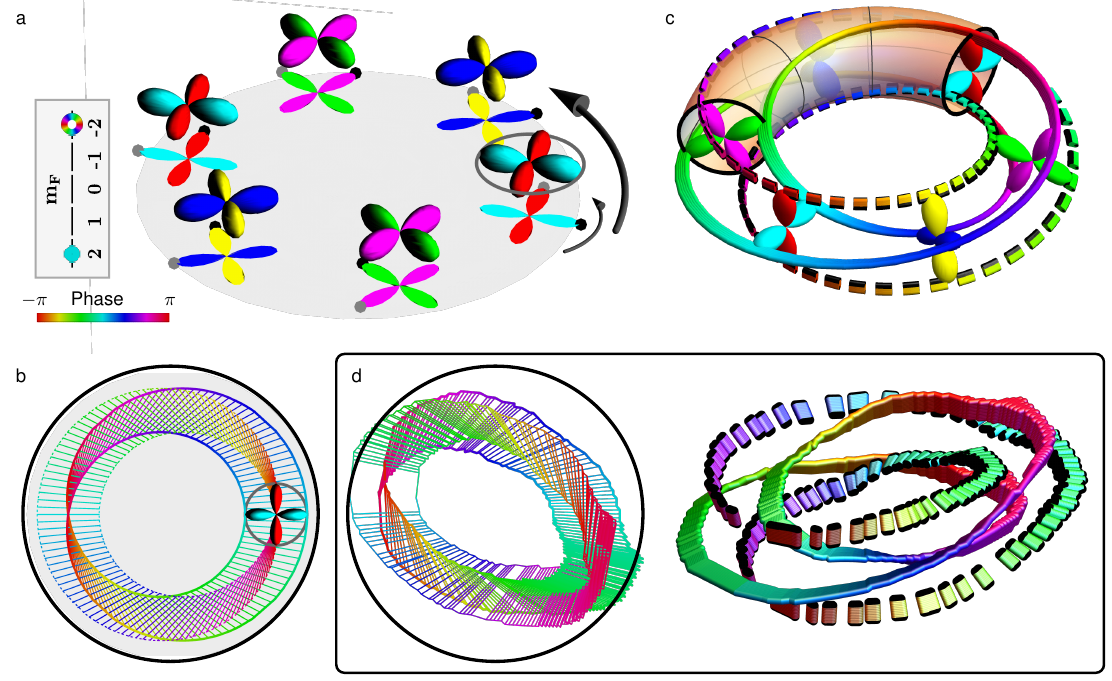}
  \caption{\textbf{Tying Solomon's knot and a discrete four-fold symmetry.} (a) We show the spherical harmonic representation of the spin-orbit invariant wavefunction $\zeta^{\text{BN}} = \begin{pmatrix}
  \zeta_2 & 0 & 0 & 0 & \zeta_{-2}e^{i2\phi} 
  \end{pmatrix}^T$ that shows a rotation by $\pi$ on an azimuthal traversal. In the inset box $m_F$ denotes the magnetic sublevels of the hyperfine states. We depict the populations and phases of the wavefunction components $|F, m_F\rangle$, showing the uniform phase of the component in $|2,2\rangle$ and the $\ell' = 2$ azimuthal phase of the component in $|2,-2\rangle$.
(b) Visualizing the orientation of the atomic wavefunction that combines a discrete four-fold symmetry with a condensate phase in terms of two disjoint M\"{o}bius-type topological structures. A pair of disjoint lobe tip paths must be constructed to fully visualize the topology of the fractional spin state rotation. (c) The two disjoint paths are represented with solid and dashed curves. The 3D representation shows that these paths are interlinked. The associated knot is the torus link $K_{2,4}$. (d) We reconstruct the local spin state and the torus link from experimental data.
  The color bar depicts phase in all panels.
  }
  \label{fig:4}
\end{figure*}

\begin{figure}[t]
  \centering

  \includegraphics[trim={0cm 0cm 0cm 0cm},clip, width=\linewidth]{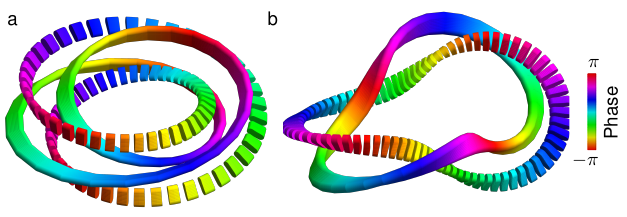}
  \caption{\textbf{Topologically equivalent knots.}  We show the knots $K_{2, 4}$ and $K_{4, 2}$, which are geometrically distinct but topologically equivalent knots differing only in the choice of meridional and longitudinal coordinate in mapping the coordinated rotation of the wavefunction onto the torus. The knot structure is a torus link that consists of two disjoint paths, represented with solid and dashed curves.
  (a) The knot $K_{2,4}$ and the topologically equivalent knot (b) $K_{4,2}$ which is Solomon's knot. 
  The color bar depicts phase in both panels.
  }
  \label{fig:5}
\end{figure}

\begin{figure}[t]
  \centering

  \includegraphics[width=\linewidth]{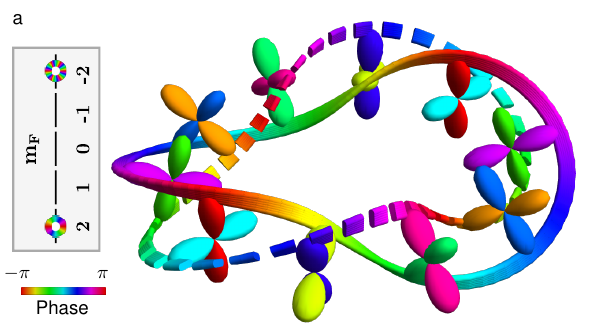}
  \caption{
 \textbf{Direct realization of Solomon's knot.} (a) We show a direct realization of Solomon's knot and a four-fold symmetry in the biaxial nematic phase 
 with $(\ell = -2, \,\ell' = 6)$ the orbital angular momentum 
 in states $|2,2\rangle$ and $|2,-2\rangle$ with a spin state rotation parameter $\lambda = 8/4$. Tracing the lobe tip paths of the spherical harmonics on an azimuthal traversal shows the appearance of the doubly-linked Solomon's knot structure $K_{4,2}$. In the inset box $m_F$ denotes the magnetic sublevels of the hyperfine states. We depict the populations and phases of the wavefunction components $|F, m_F\rangle$, showing the $\ell = -2$ phase of the component in $|2,2\rangle$ and the $\ell' = 6$ azimuthal phase of the component in $|2,-2\rangle$. The two disjoint paths that form the twice-linked Solomon's knot are represented with solid and dashed curves.
  }
  \label{fig:6}
\end{figure}

\noindent\textbf{
Solomon's knot in the biaxial nematic phase.} 
In a highlight of our work, we have created a nontrivial linked structure in the spin-2 manifold: the torus knot $K_{2,4}$, which is topologically equivalent to a $K_{4,2}$ Solomon's link. 

The spherical harmonic representation of the biaxial nematic wavefunction $\zeta^{\text{BN}}_0
\equiv\frac{1}{\sqrt{2}}\begin{pmatrix}
1 & 0 & 0 & 0 & 1
\end{pmatrix}^T$  combines a four-fold symmetry under spin state rotation around the atomic quantization axis with a condensate phase. This is the highest order of internal symmetry hosted in the spin-2 manifold. When $\ell$ and $\ell'$ are the OAM of states $|2,2\rangle$ and $|2,-2\rangle$, Eq.~\ref{eq:generalformslambdajlambda} yields $ \lambda  = (\ell'-\ell)/4$ and $j_{\lambda} = (\ell'+\ell)/2$, with associated knots $K_{\ell'-\ell,4}$. With $\ell = 0$ and $\ell' = 2$ we create a state with $(\lambda = 2/4,\,j_{\lambda} = 1)$ using $\zeta_0 = \zeta^{\text{BN}}_0$ in Eq.~\ref{eq:spingaugesymm}:
\begin{align}
 \zeta^{\text{BN}} &= \frac{1}{\sqrt{2}}\begin{pmatrix}
1\\0 \\ 0 \\ 0 \\ e^{i 2 \phi}
\end{pmatrix}\label{eq:BNVortex}.
\end{align}
A spherical harmonic visualization of this state is shown in Figure~\ref{fig:4}(a). The four-fold symmetry of the lobes of the spherical harmonics show a rotation by $\pi$ through an azimuthal traversal while the phase changes by $2\pi$. The associated torus knot is  $K_{2,4}$. In contrast to the previous two examples this is a torus link consisting of a pair of disjoint paths (shown as solid and dashed curves in the figures) that are the $K_{1,2}$ knot and its linked image. The M\"{o}bius strip topology of each is visible in a map of the directors tracking the spin wavefunction (Fig.~\ref{fig:4}(b)). The linked nature of the two paths is evident in the 3D construction (Fig.~\ref{fig:4}(c)).  The local spin state reconstructed from experimental data, and the experimentally realized knot $K_{2,4}$, are shown in Figure~\ref{fig:4}(d).  The experimental techniques for realization of this torus link are outlined in the Methods section. 

In mapping the spin rotation and the azimuthal angles onto a torus to reveal torus knots, the two directions on the torus may be interchanged, which interchanges the number of longitudinal and meridional windings. Thus, the knots $K_{2,4}$ and $K_{4,2}$ are topologically equivalent, but geometrically distinct, knots. The torus link $K_{4,2}$, known as Solomon's knot (Figure~\ref{fig:5}), displays two doubly interlinked rings and four crossings, in contrast to the simpler Hopf link. Solomon's knot $K_{4,2}$ can be more directly realised by choosing a configuration where $\ell'-\ell = 8$. Consider the specific example where $(\ell = -2,\, \ell'  = 6)$ are the OAM of $|2,2\rangle$ and $|2,-2\rangle$, so that $\lambda = 8/4$.  The doubly linked structure again appears as a pair of paths traced by the lobe tips of the spherical harmonics on an azimuthal traversal (Figure ~\ref{fig:6}). The full knot structure consists of two linked Solomon's knots $K_{4,2}$.

\noindent\textbf{\LARGE Discussion}

The topological atom optics studied here is reminiscent of knotted structures of optical polarization in monochromatic and bichromatic optical fields. Our constructions most closely map to studies in paraxial optical fields \cite{Ballantinee1501748, Pisanty2019}, where the spin and OAM of light, $\hat{S}_z$ and $\hat{L}_z$, are separable degrees of freedom. 

The symmetries of optical polarization are visualised as the path traced in time by the tip of the electric field vector in a fixed plane perpendicular to the direction of propagation. In monochromatic fields, this path is the familiar polarization ellipse. Bichromatic fields show a richer variety of polarization structures, with polarization symmetries represented by Lissajous curves that are traced when a pair of orthogonal sinusoidal signals of frequencies $(p\omega, q\omega)$ combine. Consider the field
\begin{align}\label{eq:ElectricFieldpq}
    \mathbf{E}(t) = \text{Re} \left[(E_{p +}e^{- i p \omega t}\mathbf{e}_{+} + E_{q -} e^{-i q \omega t}\mathbf{e}_{-}\right)],
\end{align} where $\mathbf{e}_{\pm} = \mp\frac{1}{\sqrt{2}}(\mathbf{e}_{ x}\pm i \mathbf{e}_{y}),\ {\rm \text{and }}  \mathbf{e}_{0} = \mathbf{e}_{z}$ relate the spherical and Cartesian bases.  $E_{p +}$ and $E_{q-}$ are complex amplitudes of field components at frequencies $p\omega$ and $q\omega$. Stable optical polarization Lissajous figures are traced for specific choices of $(p,q)$ with $p/q$ rational, for appropriate field amplitudes, and for stationary relative phase \cite{Freund2003, Kessler:03}. These can display $(p+q)$-fold discrete symmetries under spin rotation (see Supplementary Note 2).
A coupling with the optical OAM in these fields realizes spin-orbit invariant optical fields that are invariant under transformations generated by an operator of the form $\hat{J}_{\gamma} = \hat{L}_z + \gamma \hat{S}_z$, identified simply as the torus knot angular momentum, that rotates the polarization by an angle $\gamma\phi$ as the beam spatial profile is transformed through an angle $\phi$ \cite{Pisanty2019, Ballantinee1501748}. For fields with OAM $(m, m')$ associated with frequency components $(p\omega, q\omega)$ (in Eq.~\ref{eq:ElectricFieldpq}), the Lissajous figures display internal spin rotations by an angle $\gamma\phi$ on a full azimuthal traversal, with $\gamma = (m' p - m q)/(p+q)$ \cite{Pisanty2019}. This non-trivial internal polarization rotation by a fraction $2\pi\gamma$ around a closed loop realizes torus knots in the optical case \cite{Pisanty2019}. 

In the work reported here, our optical imprinting technique has allowed us to create a rich variety of stable and non-equilibrium structures in spin-1 and spin-2 atomic wavefunctions of a $^{87}$Rb spinor BEC. In the future a range of torus knots can be created in higher spin manifolds with our techniques: the spin-3 manifold for example supports torus knot $K_{\ell, 5}$ structures. 
In the future we plan to employ analytic and geometric properties of torus knots, and their relationship to braid groups and non-Abelian vortex algebra (see Supplementary Note 3), to investigate the local and global behaviours that are consequences of these knot-topological structures.

\bigbreak

\noindent \textbf{\large Methods}

\small

\noindent \textbf{Experimental details:}
We begin with a cigar-shaped $^{87}$Rb BEC prepared in a magnetic trap. The cloud is spin-polarized within the electronic ground state manifold $5^2S_{1/2}$ in state $|F, m_F\rangle = |1, -1\rangle$ or $|2, 2\rangle$. The initial BEC is released from the magnetic trap and undergoes free fall for $9\,$ms, expanding to $\approx 50\,\mu$m such that inter-atomic interactions can be neglected on the time-scales of the experiment. The Raman interaction is then applied. The optical fields including the Raman beams and the imaging beam all propagate along the long axis of the cloud, which is also the atomic quantization axis.   The experimental configuration is described in more detail in previous publications \cite{Wright:2008, Wright:2009, ARamanWaveplate, SchultzRamanFingerprints2016}.

\noindent \textbf{The Raman interaction:} 
A fundamental building block of our state preparation technique is coherent two-photon optical Raman coupling of a three-level $\Lambda$ system with ground states $|\psi_{\uparrow}\rangle$ and $|\psi_{\downarrow}\rangle$ and excited state $|e\rangle$ (Supplementary Figure 2). Raman-optical fields with Rabi frequencies $\Omega_A$ and $\Omega_B$ and phases $\phi_A$ and $\phi_B$ couple the transitions $|\psi_{\uparrow}\rangle\rightarrow |e\rangle$ and $|\psi_{\downarrow}\rangle\rightarrow |e\rangle$ and have polarizations $(\sigma_+, \pi, \sigma_-)$ allowing transitions with $\Delta m_F = (+1,0,-1)$ between ground and excited states. In order to lift the degeneracy of the spin states a small bias magnetic field $\approx11\,$Gauss is used. 
The Raman pulses are $5$--$10\,\mu$s long square pulses that can be variably detuned from the excited $F'=1$ or $F'=2$ manifolds within the $D_1$ line of $^{87}$Rb. The frequencies and temporal profiles are controlled with acousto-optic modulators. In the limit of large detuning $\Delta$ of the optical fields from the excited state, $|e\rangle$ can be adiabatically eliminated. The effective dynamics is then described by the two-level system $\{|\psi_{\uparrow}\rangle, |\psi_{\downarrow}\rangle\}$ through a unitary evolution of an initial state $|\psi_i\rangle = c_{\uparrow}|\psi_{\uparrow}\rangle + c_{\downarrow}|\psi_{\downarrow}\rangle$ to final state $|\psi_f\rangle = U|\psi_i\rangle$. For the square, diabatic pulses used in our experiment, the interaction parameters remain constant in time for the duration of the optical pulses, and the evolution can be computed simply by Eq.~\ref{eq:UanalyticRaman}. The parameters $\Omega = (|\Omega_A|^2+|\Omega_B|^2)/4\Delta$, $\alpha = \tan(|\Omega_A|/|\Omega_B|)$, $\phi = (\phi_A - \phi_B)$ are experimentally controlled parameters related to the total and relative intensities of the optical fields, and their relative phase \cite{schultzcreatingfbb_2016,SchultzRamanFingerprints2016}. 
The spatial mode and OAM  of the beams are controlled with a spiral phase plate that creates a Laguerre--Gaussian mode of charge $\ell=1$ or $\ell=2$. An interferometer is used to flip the mode handedness as necessary \cite{Wright:2008, Wright:2009, ARamanWaveplate, SchultzRamanFingerprints2016}. Coherent population transfer is achieved using rf pulses of $100$--$150\, \mu$s tuned to resonance between adjacent Zeeman sublevels.

To create the polar phase wavefunction of Eq.~\ref{eq:PolarHQVLemon}, we begin with a spin-polarized BEC in $|1,-1\rangle$, and use a pair of $(\sigma_+,\sigma_-)$ Raman beams with orbital angular momenta $\ell_A=0$ and $\ell_B=1$ to transfer atomic population to state $|1,1\rangle$, while also imprinting OAM $\ell = -1$ on the atoms. The non-rotating core ($\ell' = 0$) is in state $|1,-1\rangle$. Regions of the cloud where these two spin components have equal densities are in the polar magnetic phase.

To create the cyclic phase wavefunction of Eq.~\ref{eq:cyclicvortex}, we begin with a spin-polarized BEC in $|2,2\rangle$, and use a coherent rf transfer of atomic population from $|2,2\rangle$ to $|2,1\rangle$ followed by a multipulse Raman sequence. A pair of $(\pi$, $\sigma_-)$ Raman beams with vortex charges $(\ell_A =0, \,\ell_B =-1)$ creates a phase winding in $|2,2\rangle$, leaving a non-rotating Gaussian core in $|2,1\rangle$ (such that $(\ell = 1,\, \ell' = 0)$ are the OAM associated with spin states $|2,2\rangle$ and $|2,1\rangle$). A second $(\sigma_+, \,\sigma_-)$ Gaussian Raman pulse pair transfers the non-rotating core from $|2,1\rangle$ to $|2,-1\rangle$. In the cyclic magnetic phase the two spin components $|2,-1\rangle$ and $|2,2\rangle$ satisfy $|\zeta_{-1}|/|\zeta_{2}|=\sqrt{2}$.

To create the biaxial-nematic phase wavefunction of Eq.~\ref{eq:BNVortex}, we begin with atomic population in $|2,2\rangle$. A single pulse pair of $(\sigma_-,\sigma_+)$ Raman beams with vortex charges $(\ell_A=1, \ell_B=0)$ and experimental parameters favoring a four-photon process transfers atomic population to $|2,-2\rangle$. This is accomplished by tuning the two-photon transfer from $|2,2\rangle$ to $|2,0\rangle$ off resonance. The transferred atomic population picks up two units of OAM leaving behind a non-rotating core, thus creating the target spin-orbit invariant state $(\ell=0, \ell'=2)$.

\noindent \textbf{Spherical-harmonics representation:}
To visualize the symmetries of the spinor wavefunction we show the surface of  $|Z(\theta,\phi)|^2$ and colour indicating $\arg \left[Z(\theta,\phi)\right]$ where 
\begin{align}
Z(\theta,\phi) = \sum\limits_{m= -F}^F\zeta_{m}Y_{F,m}(\theta,\phi)
\end{align} 
expands the spinor $ \zeta = (\zeta_{F}, \zeta_{F-1},...,\zeta_{-F})^T$ in terms of the spherical harmonics $Y_{F,m}(\theta,\phi)$, such that $(\theta,\phi)$ define the local spinor orientation in spherical coordinates and colour represents the phase.

\noindent \textbf{Imaging:} 
In our time-of-flight Stern--Gerlach absorption imaging process an inhomogeneous magnetic field is briefly pulsed on $13\,$ms after the cloud is released, followed by a time of flight $t_f\approx 13\,$ms. A resonant, collimated, imaging beam illuminates the cloud, casting an absorption shadow in the transmitted beam which is then imaged on a CCD camera. 

Stern--Gerlach imaging gives spatially resolved density information within each spin component. The donut-shaped intensity profiles of the Laguerre--Gaussian beams result in the spatially-dependent population transfer, determining radially-dependent magnetic phases (i.e., we achieve different population in the $m_F$ levels that remain constant during the expansion of the atom cloud and characterize the local magnetic phase). The spherical harmonics are reconstructed using the imaged spin state amplitudes.

The relative phases of the Raman fields are imprinted onto the spin states by the Raman transfer, thus defining the relative phases of the atomic spin states after transfer. In particular, we have shown experimentally how the phases of the Raman fields and hence the OAM from the optical beams is transferred onto the spin states of the BEC with this  process  by using matter-wave interferometry \cite{Wright:2009, Wright:2008}. The relative phases of the atomic spin states are therefore experimentally known from our calibration of the Raman beam OAM transfer for each of the data sets shown here, and are used to reconstruct the orientation of the local spin state of the cloud (see Supplementary Note 4 and Supplementary Note 5)
.

For the $n$-fold symmetry, we construct the 2D torus knot structures by following the $n$ lobe tip paths of spherical harmonics on a full $2\pi$ traversal of the azimuthal coordinate. Both coordinates  $\phi$, the azimuthal angle, and $\tilde{\phi}$, the angle of spin rotation (as defined in the Torus knot topology section), define angles in the 2D space transverse to the quantization axis (Figure~\ref{fig:1}(a)).  To represent these structures in 3D, we use a mapping that re-orients the coordinate $\tilde{\phi}$ at each azimuth to define the meridional direction of a 3D torus, and then the azimuthal coordinate $\phi$ defines the longitudinal direction (Figure~\ref{fig:1}(b)). This is equivalent to performing rotations of the local spherical harmonics: first a rotation by $\pi/2$ about the $x$ axis and then a rotation by $\phi$ about the $z$ axis. The $n$ lobe tips of the spherical harmonics then describe structures in 3D. 

\renewcommand{\thefigure}{S\arabic{figure}}
\setcounter{figure}{0}    

\noindent \textbf{Data availability} The data used in this work are publicly archived in the Zenodo repository at \href{https://doi.org/10.5281/zenodo.8239421}{https://doi.org/10.5281/zenodo.8239421}.

\noindent \textbf{Code availability}
The Mathematica 12 notebooks used in this work are publicly archived in the Zenodo repository at \href{https://doi.org/10.5281/zenodo.8239421}{https://doi.org/10.5281/zenodo.8239421}.
\bibliographystyle{naturemag_no_url}
\bibliography{Maitreyi_Updated}

\noindent \textbf{Acknowledgements}
We thank Azure Hansen and L. S. Leslie for contributing to the data we use to show the knotted structures. This work was supported by NSF grant PHY 1708008 and NASA/JPL 
through RSAs including 1656126.

\noindent \textbf{Author contributions}
M. Jayaseelan was the primary contributor to both the experimental work and the conception of the torus knot mapping and analysis.  J. D. Murphree and J. T. Schultz participated in many of the experiments reported here.  J. Ruostekoski made key contributions to the theory and interpretation.  N. P. Bigelow was the PI and overall lead of the team both in experiment and theory.

\noindent \textbf{Competing interests} The authors declare no competing interests.

\end{document}


\title{Supplementary Information: Topological atom optics and beyond with knotted quantum wavefunctions}
\author{Maitreyi Jayaseelan}
\affiliation{Department of Physics and Astronomy\char`,\, University of Rochester\char`, \,Rochester\char`,\, NY 14627\char`,\, USA}
\affiliation{Center for Coherence and Quantum Optics\char`,\,University of Rochester\char`,\, Rochester\char`,\, NY 14627\char`, \,USA}	  

\author{Joseph D. Murphree}
\affiliation{Department of Physics and Astronomy\char`,\, University of Rochester\char`, \,Rochester\char`,\, NY 14627\char`,\, USA}
\affiliation{Center for Coherence and Quantum Optics\char`,\,University of Rochester\char`,\, Rochester\char`,\, NY 14627\char`, \,USA}	

\author{Justin T. Schultz}
\affiliation{Center for Coherence and Quantum Optics\char`,\,University of Rochester\char`,\, Rochester\char`,\, NY 14627\char`, \,USA}	
\affiliation{The Institute of Optics\char`, \,University of Rochester\char`, \,Rochester\char`, \,NY 14627\char`,\, USA}

\author{Janne Ruostekoski}
\affiliation{Department of Physics\char`,\,Lancaster University\char`,\,Lancaster\char`,\,LA1 4YB\char`,\,United Kingdom}

\author{Nicholas P. Bigelow}
\affiliation{Department of Physics and Astronomy\char`,\, University of Rochester\char`, \,Rochester\char`,\, NY 14627\char`,\, USA}
\affiliation{Center for Coherence and Quantum Optics\char`,\,University of Rochester\char`,\, Rochester\char`,\, NY 14627\char`, \,USA}	
\affiliation{The Institute of Optics\char`, \,University of Rochester\char`, \,Rochester\char`, \,NY 14627\char`,\, USA}
\date{\today}
	\maketitle

\renewcommand{\thesection}{\MakeUppercase{Supplementary Note }\arabic{section}}

\section{Magnetic phases and discrete symmetries of spinor wavefunctions}	

\makeatletter
\def\fnum@figure{\figurename\nobreakspace\textbf{\thefigure}}
\makeatother

\renewcommand{\figurename}{\textbf{Supplementary Figure}}

\begin{figure}[ht]
  \centering
  \includegraphics[width=0.2\linewidth]{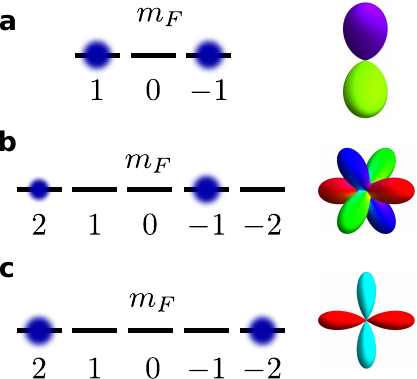}
  \caption{\textbf{Magnetic phases and discrete symmetries of spinor wavefunctions.} Spherical-harmonic representation of spinors combining discrete symmetries and condensate phase, (a) spin-1 polar phase, (b) spin-2 cyclic phase,  (c) spin-2 biaxial nematic phase. Here $m_F$ denotes the magnetic sublevels of the hyperfine states.
}
\label{fig:S1}
\end{figure}

For weak Zeeman shifts, the magnetic phases can be defined as dynamically stable stationary solutions of the mean-field theory from their condensate-spin magnitude and spin-singlet amplitudes~\cite{KawaguchiUedaSpinorBoseEinsteinCondensates}. The condensate-spin expectation value $\langle\Fhat\rangle=\sum_{\alpha\beta} \zeta_\alpha^\dagger\Fhat_{\alpha\beta} \zeta_\beta$ is obtained from
 the vector $\Fhat$ of spin-1 or spin-2 matrices. 
 
For the spin-1 polar magnetic phase (Supplementary 
Figure~\ref{fig:S1}(a)) the spin magnitude vanishes $|\langle\Fhat\rangle|=0$ and the corresponding order-parameter symmetry $(S^1\times S^2)/\mathbb{Z}_2$ can be described by the condensate phase $\tau$ and an unoriented nematic axis $\hat{\bf d}$, such that the general polar state can be expressed as $\frac{\exp(i\tau)}{\sqrt{2}}\begin{pmatrix}
 -d_x+i d_y, & \sqrt{2} d_z, & d_x+i d_y
\end{pmatrix}^T$. The nematic axis can also be specified for a general spin-1 wave function~\cite{Lovegrove16}. The polar magnetic phase state displays nematic order. For example, the spinor $\zeta^{\text{P}}_0 = \frac{1}{\sqrt{2}}\begin{pmatrix}
 1 & 0 & 1
\end{pmatrix}^T$ can display a two-fold symmetry under spin rotations around the $z$ axis when compensated by the condensate phase.
 
The spin-2 magnetic phases are characterized by the values of $|\langle\Fhat\rangle|$, $|A_{20}|$, and $|A_{30}|$ , where
\begin{align}
    A_{20}&=\frac{1}{\sqrt{5}}(2\zeta_{+2}\zeta_{-2}-2\zeta_{+1}\zeta_{-1}+\zeta_0^2),\\
    A_{30}&=\frac{3\sqrt{6}}{2}(\zeta_{+1}^2\zeta_{-2}+ \zeta_{-1}^2\zeta_{+2}) + \zeta_0 (
-6\zeta_{+2}\zeta_{-2} -3\zeta_{+1}\zeta_{-1}+\zeta_0^2),
\end{align} 
are the amplitudes of spin-singlet pair and  spin-singlet trio formation, respectively. 

The spin-2 cyclic magnetic phase (Supplementary 
Figure~\ref{fig:S1}(b)) has $|\langle\Fhat\rangle|=|A_{20}|=0$. The spinor $\zeta^{\text{C}}_0
=\frac{1}{\sqrt{3}}\begin{pmatrix}
1 & 0 & 0 & \sqrt{2} & 0
\end{pmatrix}^T$, and the analogous alternative $\frac{1}{\sqrt{3}}\begin{pmatrix}
0 &  \sqrt{2} & 0 & 0 & 1
\end{pmatrix}^T$, combines the condensate phase with a three-fold symmetry about the $z$ axis in which case the three lobes of the spherical-harmonics  are out of phase by $2\pi/3$. A four-fold symmetry in the shape of the spherical harmonics about the $z$ axis is obtained with a different orientation of the order parameter, as for the spinor $\frac{1}{2}\begin{pmatrix}
1 & 0 & i\sqrt{2} & 0 & 1
\end{pmatrix}^T$.

The spin-2 biaxial nematic phase (Supplementary 
Figure~\ref{fig:S1}(c)) has  $|\langle\Fhat\rangle|=0$ and
$|A_{20}|=\frac{1}{\sqrt{5}}$. It can be distinguished from the uniaxial case by  the value $|A_{30}|=0$. The spinor $\zeta^{\text{BN}}_0
=\frac{1}{\sqrt{2}}\begin{pmatrix}
1 & 0 & 0 & 0 & 1
\end{pmatrix}^T$ combines the condensate phase with a four-fold symmetry about the $z$ axis.

\begin{figure}[ht]
  \centering
 \includegraphics[width=0.3\linewidth]{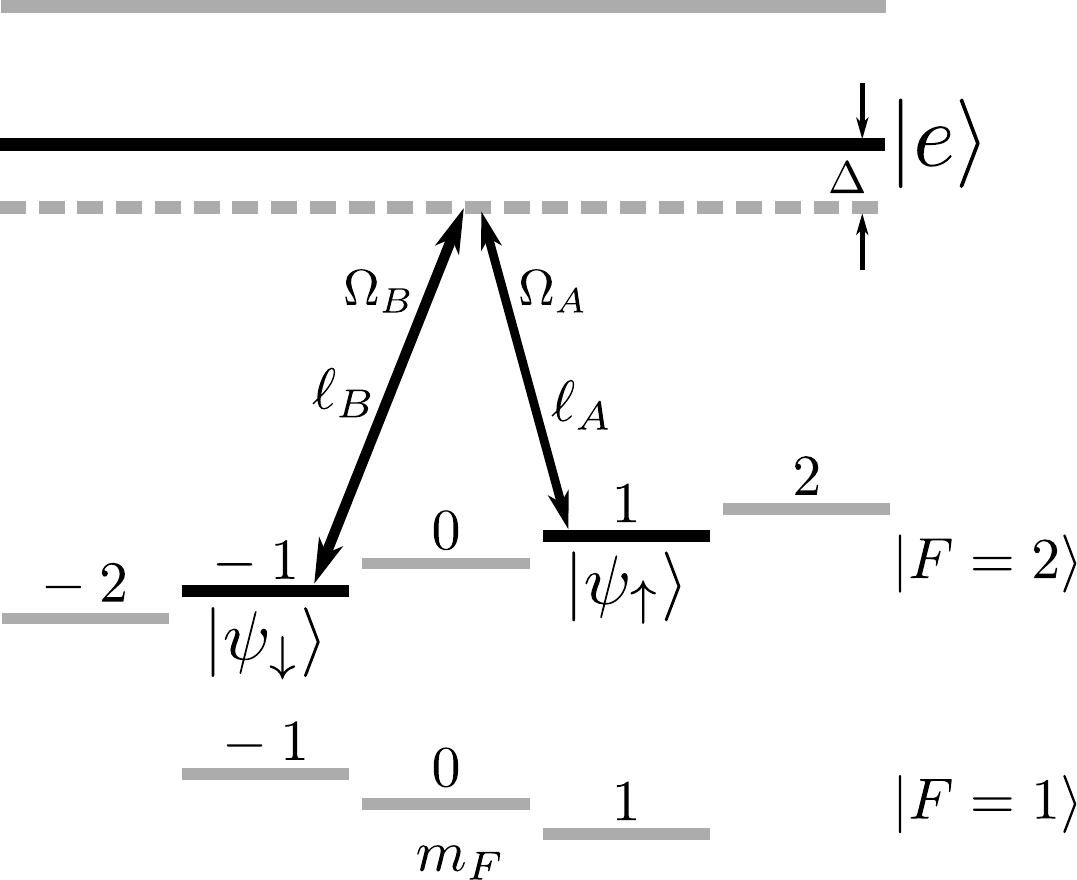}
  \caption{\textbf{Raman coupling with optical fields.} Two-photon Raman coupling between two atomic Zeeman sublevels within a spin-$F$ hyperfine manifold. Here $m_F$ denotes the magnetic sublevels of the hyperfine states $|F=1\rangle$ and $|F=2\rangle$. Two optical fields with Rabi frequencies $\Omega_A$ and $\Omega_B$, and orbital angular momentum $\ell_A$ and $\ell_B$ couple two Zeeman sublevels $|\psi_{\uparrow}\rangle$ and $|\psi_{\downarrow}\rangle$ via the excited state $|e\rangle$, with single-photon detuning $\Delta$.
 }

\label{fig:S2}
\end{figure}

In each case, a discrete spin rotation by $2\pi/n$ where $n$ is the degree of internal symmetry leaves the spinor unchanged up to a phase. Supplementary Figure~\ref{fig:S2} shows Raman coupling with optical fields. Different configurations of fields as outlined in the Methods section (main text) realize different spin-orbit invariant states.

\section{Comparison with the optical polarization M\"{o}bius strip and torus knot structures}

\begin{figure*}[t]
  \centering
  \includegraphics[width=0.7\linewidth]{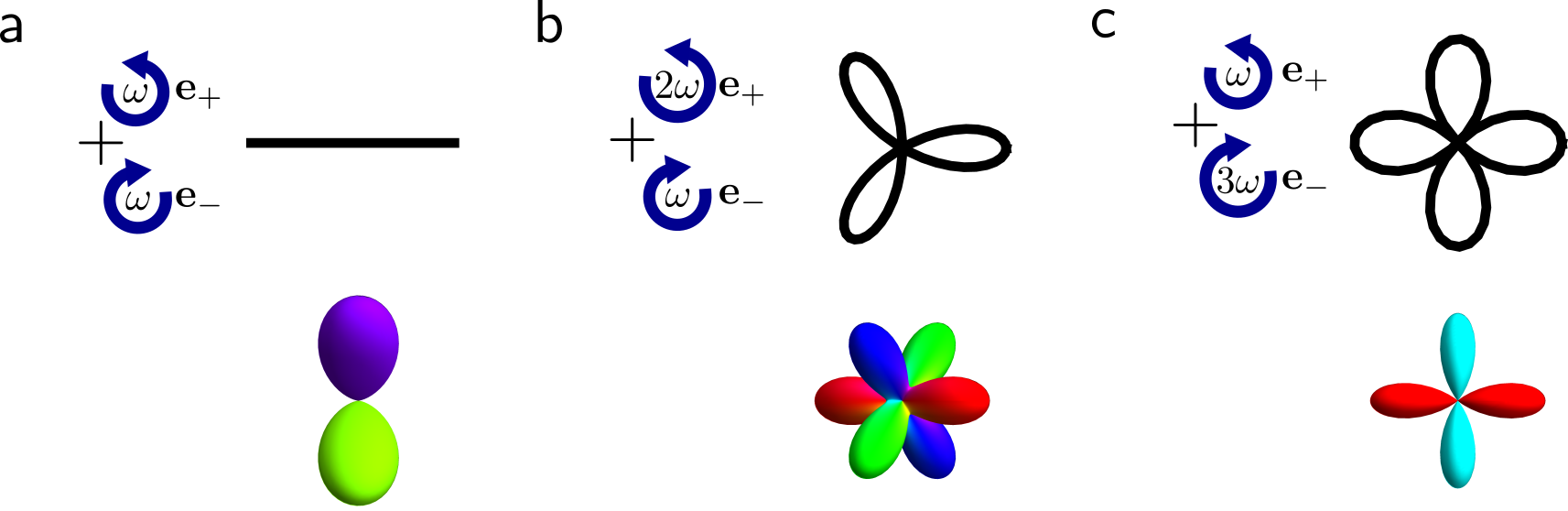}
  \caption{\textbf{Comparison of optical and atomic cases.} Atomic wavefunctions with discrete spin rotational symmetries and corresponding optical polarization Lissajous figures with the same symmetries. (a) showcases two-fold symmetries. The polarization figure of monochromatic optical fields with frequency $\omega$ and both circular polarization components $\mathbf{e}_{+}$ and $\mathbf{e}_{-}$ is a line. The atomic analog is the polar phase wavefunction with two-fold discrete symmetry, up to a phase. (b) shows three-fold symmetries. A three-fold symmetric polarization figure is obtained in the case of a bichromatic optical field with frequency components  $\omega$ and $2\omega$ in circularly polarized states $\mathbf{e}_{+}$ and $\mathbf{e}_{-}$. The atomic analog is the cyclic phase wavefunction with three-fold discrete symmetry, up to a phase. (c) shows four-fold symmetries. Polarization figures of four-fold symmetry are obtained in the case of bichromatic optical field with frequency components  $\omega$ and $3\omega$ in circularly polarized states $\mathbf{e}_{+}$ and $\mathbf{e}_{-}$. The atomic analog of the biaxial nematic phase wavefunction combines discrete four-fold symmetry with the condensate phase. 
  }
  \label{fig:S6}
\end{figure*}

The spin-orbit invariant atomic wavefunctions we realise here are analogous to coordinated rotation invariant optical fields as described in the main text and Supplementary Refs.~\cite{Pisanty2019, Ballantinee1501748,Bauer2015OpticalMobius, BauerMultiTwistpolarizationRibbon:2019}. 
Supplementary Figure~\ref{fig:S6} shows examples of the spherical harmonic representations of the spin-1 and spin-2 atomic wavefunctions with two-, three-, and four-fold symmetries, up to the condensate phase, and corresponding optical polarization Lissajous figures that display similar symmetries. 

\renewcommand{\thesubsection}{\MakeUppercase{S-}\arabic{section} \Alph{subsection}}

\subsection{Spin-1}
A direct correspondence between paraxial monochromatic optical fields and spin-1 atomic systems is available from angular momentum considerations, and is established through the mapping 
\begin{equation}
\mathbf{e}_{\pm}\mapsto |1,\pm1\rangle ,\quad
\mathbf{e}_0 \mapsto |1,0\rangle.\nonumber\label{eq:spin1mapping}
\end{equation} 
The electric field analogous to the spin-1 wavefunction  
$\zeta = \begin{pmatrix}
 \zeta_1 e^{i\ell\phi} & 0 & \zeta_{-1}e^{i\ell'\phi}
\end{pmatrix}^T$
is thus the monochromatic optical field:
\begin{align}\label{eq:opticsMobiusArbitm}
    \mathbf{E}(t) = \text{Re} \left[\left(E_{+}e^{i m \phi}\mathbf{e}_{+} + E_{-}e^{i m'\phi} \mathbf{e}_{-}\right) e^{-i\omega t}\right],
\end{align}
where orbital angular momenta $m$ and $m'$ are coupled to the circularly polarized states $\mathbf{e}_{+}$ and $\mathbf{e}_{-}$. The correspondence between the same symmetries of the atomic and optical cases for the amplitudes reads as $|\zeta_{1}|\mapsto |E_+|$ and $|\zeta_{-1}|\mapsto |E_-|$ and  for the phase windings as $\ell\mapsto m$ and $\ell'\mapsto m'$. Purely two-fold symmetry is obtained when $|\zeta_1| = |\zeta_{-1}|$; this state corresponds to linear polarization in optics.

\subsection{Spin-2}
Optical polarization representations that exhibit three- and four-fold symmetries studied in the atomic case can be realised in bichromatic optical fields.
Trefoil electric fields analogous to the cyclic phase wavefunction  $\zeta = \begin{pmatrix}
 \zeta_2 e^{i\ell\phi} & 0 & 0 & \zeta_{-1}e^{i\ell'\phi} & 0
\end{pmatrix}^T$ are found in the bichromatic field with frequencies $(\omega, 2\omega)$:
\begin{align}\label{eq:trefoil}
    \mathbf{E}(t) = \text{Re}\left[E_{2+} e^{i m\phi}e^{-i2\omega t}\mathbf{e}_{+} + E_{1-} e^{i m'\phi}e^{-i\omega t}\mathbf{e}_{-}\right]
\end{align} where $m$ and $m'$ are the orbital angular momenta associated with the polarizations $\mathbf{e}_{+}$ and $\mathbf{e}_{-}$ and frequencies $2\omega$ and $\omega$. It is useful to recognize the correspondence between the evolution frequencies of the spin states and the optical frequencies. This correspondence allows an easier identification of atomic states with discrete symmetries and corresponding optical fields carrying the same symmetries.

The optical three-fold symmetric case is obtained when $|E_{2+}| = |E_{1-}|$, as opposed to the atomic example of $|\zeta_{-1}| = \sqrt{2}|\zeta_2|$, representing the correspondence $|\zeta_{-1}|\mapsto \sqrt{2}|E_{1-}|$ and $|\zeta_{2}|\mapsto |E_{2+}|$. Moreover, the polarization rotation parameter $\gamma$ for the trefoil optical field takes the value \cite{Pisanty2019}
\begin{align}
    \gamma = (2m'-m)/3 \label{eq:trefoilGamma},
\end{align} in contrast to the corresponding atomic parameter $\lambda = (\ell'-\ell)/3$ (Eq.2 of the main text). We recognize these differences as stemming from the spin-1 nature of photons.

Finally, the examples with the biaxial nematic phase that combines discrete four-fold symmetries with the condensate phase  find analogs in bichromatic optical fields with frequencies $(\omega, 3\omega)$. Here a comparison between the atomic spin frequencies and the optical field frequencies no longer reproduces the same symmetries, since the spin evolution frequencies $(2\omega_L, -2\omega_L)$ are not coprime. A closer analogy to the bichromatic optical field with frequencies $(\omega, 3\omega)$ can be found in a spin-3 manifold.

\section{Connections to Defects}

Vortex classification based on the fundamental homotopy group is a well developed theory that is accepted as a complete description, including, e.g., the description of non-Abelian vortex algebra. However, the fundamental homotopy group description is also rather abstract. 
Here we consider physically intuitive characterizations of line defects (singular vortices) based on the geometry of the magnetic phase manifolds in a way that follows directly from the knot construction. 

In our treatment the symmetries of the spin-2 magnetic phases are described within the Majorana representation~\cite{majorana_nuovocimento_1932,Barnett2006} (or by the shape of the lobes of the spherical-harmonics representation). For example, the cyclic phase exhibits a discrete internal Majorana symmetry of a tetrahedron. In this framework line defects and vortices follow from Eq.~(1) in the text, where the macroscopic phase rotations and spin rotations $T^2=S^1\times S^1$ are coordinated along the symmetry axes. As discussed on page 2, according to the single-valuedness of the wavefunction, the rotations are related by $m/n$, where $m$ is the number of fractional spin rotations through angle $2\pi /n$ of the wavefunction around an $n$-fold symmetry axis, resulting in a $K_{m,n}$ knot. 

Every line defect conjugacy class then follows from the analysis of $K_{m,n}$ along each symmetry axis of a tetrahedron. In addition to the trivial line defects of full $2\pi$ phase and spin rotations, we have $K_{1,2}$ defects from the rotations around the axis joining the centers of the edges to the center of the tetrahedron, $K_{1,3}$ and $K_{-1,3}$ conjugacy classes from the $2\pi/3$ rotations around the lines joining the vertices or the center points of the faces of the tetrahedron to the center the tetrahedron, and  $K_{\pm 2,3}$ for the analogous $4\pi/3$ rotations. The commutation relations follow from the conjugacy classes.

\section{Reconstructing the local wavefunction from experimental data}

To reconstruct the torus knot structures from our experimental data, we identify the portions of the cloud where the atomic populations are in the required ratio to realise specific discrete symmetries. The raw data is magnified by four times and blurred with a Gaussian blur. This data is fit to a fitting function that is the sum of Gaussian and Laguerre--Gaussian intensity profiles
\begin{align}
    |G(x,y)|^2 & = \frac{1}{2 \pi \sigma_x\sigma_y}\exp{\left(-\frac{(x-x_{0g})^2}{2\sigma_x^2}-\frac{(y-y_{0g})^2}{2\sigma_y^2}\right)} \\
    |LG(x,y)|^2 & = \frac{4}{\pi w_x w_y}\exp{\left(-2\left(\frac{(x-x_{0lg})^2}{w_x^2}+\frac{(y-y_{0lg})^2}{w_y^2}\right)\right)\left(\frac{(x-x_{0lg})^2}{w_x^2}+\frac{(y-y_{0lg})^2}{w_y^2}\right)},
\end{align} weighted to preserve normalization: 
\begin{align}
    I(x,y) & = a |G(x,y)|^2 + (1-a) |LG(x,y)|^2.
\end{align} Here $(x_{0g}, y_{0g})$ and  $(x_{0lg}, y_{0lg})$ are the centers of the Gaussian and Laguerre--Gaussian intensity profiles,  $(\sigma_x, \sigma_y)$ and $(w_x, w_y)$ are the beam waists, and $a$ a weighting factor. The absorption images of atomic population in each spin state are superposed so that the mode centers coincide. The atomic spin state is then determined at each pixel based on the relative populations between the spin states, by normalizing the total population distributed among the spin states in each pixel.

\section{Cyclic phase: effect of residual population in unwanted spin states}
\begin{figure*}[t]
  \centering
  \includegraphics[width=0.5\linewidth]{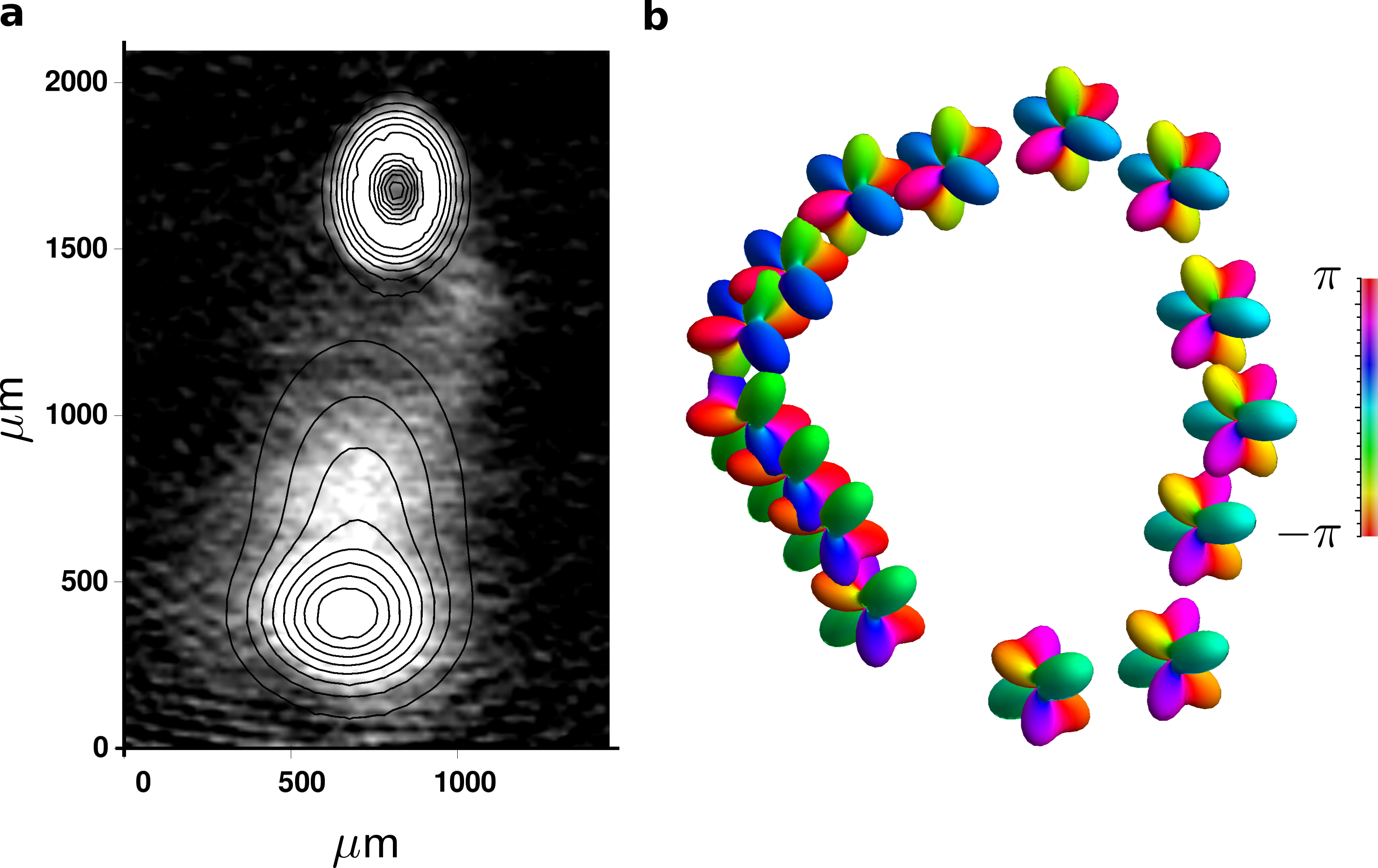}
  \caption{\textbf{Effect of residual population in unwanted spin states.} We show the effect of residual population in unwanted spin states on the cyclic phase wavefunction. (a) Spin-orbit invariant cyclic phase wavefunction of the main text   
  $\zeta^{\text{C}} = \begin{pmatrix}
  \zeta_2 e^{i \phi} &  0 & 0 & \zeta_{-1} & 0 
  \end{pmatrix}^T$ shows residual population in an unwanted spin state $|2,0\rangle$. (b) Spherical harmonics reconstruct the spin-orbit invariant state while taking this spurious population into account. The three-fold symmetry is still evident.
  }
  \label{fig:S3}
\end{figure*}

Reconstruction of the cyclic phase wavefunctions are shown in Fig. 3 of the main text, using populations in the relevant spin states $|2,2\rangle$ and $|2,-1\rangle$. The experimental data shows a small residual population in state $|2, 0\rangle$ that was neglected in this reconstruction. Here we examine the impact of this residual population on the knot topology (Supplementary 
Figure~\ref{fig:S3}).

We fit the populations of the spin states to a function that is the sum of a Laguerre--Gaussian intensity profile and two Gaussians with different centers and widths. The integrated population in the state $|2,0\rangle$ is found to be $\approx 23\%$. We reconstruct the spherical harmonics describing the local spin state, now taking the atomic population in $|2,0\rangle$ into account (the phase is assumed to be the same as that of the Gaussian portion of the cloud in $|2,-1\rangle$). The spurious population is seen to contribute some asymmetry to the spherical harmonics but the three-fold symmetry remains robust. 

The residual population in $|2,0\rangle$ is a consequence of our state preparation technique for this wavefunction.
We use a resonant rf field to transfer atomic population from $|2,2\rangle$ to $|2,1\rangle$ before the multipulse Raman sequence. For the magnetic fields and pulse lengths and powers used to effect a complete transfer of atomic population out of $|2,2\rangle$, some of the population is driven to the next adjacent state $|2,0\rangle$. This population can be minimized by using a different set of parameters to effect the population transfer. It could further be completely eliminated by using a resonant optical pulse to selectively deplete population in $|2,0\rangle$ either before or after the Raman sequences while leaving the other spin state populations unaffected.

\renewcommand\refname{Supplementary References}
\bibliographystyle{nature_no_url_copy}
\bibliography{Maitreyi_Updated2}